# Structural Phase Transitions in SrTiO$_3$ from Deep Potential Molecular Dynamics

Ri He[1#], Hongyu Wu[1#], Linfeng Zhang[2,3], Xiaoxu Wang[2,3], Fangjia Fu[3,4], Shi Liu[5,6,7*] and Zhicheng Zhong[1,8*]

[1]Key Laboratory of Magnetic Materials Devices & Zhejiang Province Key Laboratory of Magnetic Materials and Application Technology, Ningbo Institute of Materials Technology and Engineering, Chinese Academy of Sciences, Ningbo 315201, China

[2]DP Technology, Beijing 100080, China

[3]AI for Science Institute, Beijing 100080, China

[4]School of mathematical sciences, Peking University, Beijing 100871, China

[5]School of Science, Westlake University, Hangzhou, Zhejiang 310024, China

[6]Institute of Natural Sciences, Westlake Institute for Advanced Study, Hangzhou, Zhejiang 310024, China

[7]Key Laboratory for Quantum Materials of Zhejiang Province, Hangzhou Zhejiang 310024, China

[8]China Center of Materials Science and Optoelectronics Engineering, University of Chinese Academy of Sciences, Beijing 100049, China

# Abstract

Strontium titanate (SrTiO$_3$) is regarded as an essential material for oxide electronics. One of its many remarkable features is subtle structural phase transition, driven by antiferrodistortive lattice mode, from a high-temperature cubic phase to a low-temperature tetragonal phase. Classical molecular dynamics (MD) simulation is an efficient technique to reveal atomistic features of phase transition, but its application is often limited by the accuracy of empirical interatomic potentials. Here, we develop an accurate deep potential (DP) model of SrTiO$_3$ based on a machine learning method using data from first-principles density functional theory (DFT) calculations. The DP model has DFT-level accuracy, capable of performing efficient MD simulations and accurate property predictions. Using the DP model, we investigate the temperature-driven cubic-to-tetragonal phase transition and construct the in-plane biaxial strain-temperature phase diagram of SrTiO$_3$. The simulations demonstrate that strain-induced ferroelectric phase is characterized by two order parameters, ferroelectric distortion and antiferrodistortion, and the ferroelectric phase transition has both displacive and order-disorder characters. This works lays the foundation for the development of accurate DP models of other complex perovskite materials.

---

[#] These authors contribute equally to this work
[*] liushi@westlake.edu.cn
[*] zhong@nimte.ac.cn

# I. INTRODUCTION

Perovskite oxides with formula $ABO_3$ is an important class of functional materials. In the $ABO_3$ perovskite structure, the A site cation is twelve-fold coordinated and the B site cation is six-fold coordinated with oxygen anions, respectively. Many different cations and anions can be incorporated in this structure, resulting in a great number of functional materials that hold great promises for new electronic device applications [1-3]. Strontium titanate ($SrTiO_3$, STO) as a representative $ABO_3$ perovskite oxide has long captured the attention of condensed matter physicists, because of its rich physical properties such as two-dimensional electron gas [4-6], Rashba spin-orbit coupling[7-9], interfacial superconductivity[10,11] and multiple structural instabilities[12]. Among them, the incipient ferroelectricity is a unique structural property of STO that has attracted extensive studies [13-15]. At room temperatures, STO has a cubic structure with *Pm-3m* space group. Below 105 K, it undergoes an antiferrodistortive (AFD) structural transition, transforming to a tetragonal *I4/mcm* phase. The AFD structure can be described as neighboring $TiO_6$ octahedra rotating a small angle in the opposite directions around [001] axis. The tetragonal phase is paraelectric (i.e. incipient ferroelectricity), where the quantum fluctuations and AFD instabilities suppress long-range polarization ordering. Despite the apparent paraelectric nature of STO, it exhibits many abnormal phenomena indicating its proximity to a ferroelectric phase [13,14]. Many experiments have demonstrated that the ferroelectric phase can be induced via various methods, including epitaxial strain from substrate [16,17], defect engineering [18], dimension reduction[19], cation doping[20], and optical and terahertz electric field excitation [21,22].

The tetragonal-to-cubic phase transition in STO has been intensively studied with thermodynamic Landau-Ginsburg-Devonshire (LGD) model [23-26]. Nevertheless, the LGD model is often subjected to two well-known limitations: *i*). the material coefficients involved in the Landau free energy expansion are difficult to determine, and a slight variance of coefficient value may cause a drastic change of the resulting phase diagram [25]; *ii*). such phenomenological models cannot reveal the mechanisms of phase transition at the atomic level. First-principles density functional theory (DFT) calculations can accurately predict the atomic structures of cubic and tetragonal phases of STO and their small energy difference (~1.1 meV/atom) as well as the quantitative relationship between strain and polarization [27,28], but the expensive computational cost prevents the application of DFT from studying STO phase transition over a larger length and/or time scale (e.g. limited to a system of hundreds of atoms in the time scale of pico-second). Classical molecular dynamics (MD) simulation is a very effective technique for understanding the structural phase transition mechanisms. Several empirical interatomic potentials have been developed to simulate the cubic structure of STO [29-31]. However, none of them has been used to explore the cubic-to-tetragonal phase transition, likely because the subtle energy variation (~ 1 meV/atom) induced by the structural transition is beyond the energy resolution of the analytical function used to construct the interatomic potential [29]. Therefore, accurate MD simulations of phase transition in STO demands a high-accuracy potential.

In recent years, machine learning methods have been used as a powerful tool to develop the interatomic potential of crystalline and molecular materials. For instance, the Behler–Parrinello neural network [32], the Gaussian approximation potentials [33], the Select Neighbors and Parameters[34] and the SchNet [35] have achieved great success. Among them, the recently proposed Deep Potential (DP) model based on deep neural network can provide a DFT-level accurate interatomic potential by training from a large dataset of DFT energies and forces of a wide range of atomic configurations [36,37]. The well-trained deep-neural- network potential can be used to perform MD simulations without sacrificing the accuracy. In recent works, it has been demonstrated that the energy error of a DP model relative to DFT is within milli-electron-volt per atom for the ground-state configuration [38-42]. Whether the DP model could describe the subtle free energy variation of just ~1.1 meV/atom during tetragonal to cubic phase transition in STO remains a question.

In this work, we develop an accurate DP model for STO base on a deep learning method trained on the data from DFT calculations. The accuracy of the DP models is examined by comparing with DFT results. The temperature-driven tetragonal to cubic phase transition of STO is well captured by deep potential MD (DPMD) simulations. Finally, the ferroelectric-paraelectric phase transition temperatures at different in-plane strains are calculated by the DP model, and the nature of ferroelectric phase transition is characterized at the atomic level. These DPMD simulation results provide atomistic insights into the microscopic ferroelectricity mechanism of STO under strain.

## II. COMPUTATIONAL METHODS

### A. Deep learning of interatomic potential of SrTiO$_3$

In DP model, the potential energy $E$ of a configuration is assumed to be a sum of each atomic energy $E_i$ of atom $i$, which is mapped from a descriptor $D_i$ through an embedding network. The descriptor $D_i$ characterizes the local environment of atom $i$ within a cutoff radius $R_c$. Here, the $R_c$ is set to 6 Å. The maximum number of atoms within the $R_c$ is set to 160 for Sr, 160 for Ti, and 480 for O, respectively. The translational, rotational, and permutational symmetry of the $D_i$ are preserved by an embedding network. The smooth edition of deep potential model was employed to remove the discontinuity introduced by the cutoff radius [37]. The size of embedding network and fitting network is (25, 50, 100) and (240, 240, 240), respectively. Residual network was employed in the training of fitting network. The loss function is defined as:

$$L(p_\varepsilon, p_f, p_\xi) = p_\varepsilon \Delta \epsilon^2 + \frac{p_f}{3N}\sum_i |\Delta \boldsymbol{F}_i|^2 + \frac{p_\xi}{9}\|\Delta \xi\|^2, (1)$$

where, $\Delta \epsilon$, $\Delta \boldsymbol{F}_i$, $\Delta \xi$ represent the energy, force, and virial tensor difference between the DP model prediction and the training dataset (come from DFT calculations). The $p_\varepsilon$, $p_f$, and $p_\xi$ are weight coefficient of energy, force, and virial tensor. The $p_\varepsilon$ and $p_\xi$ increase from 0.02 to 1 and $p_f$ decreases from 1000 to 1 during the training procedure. The DP models are trained with 1200000 steps.

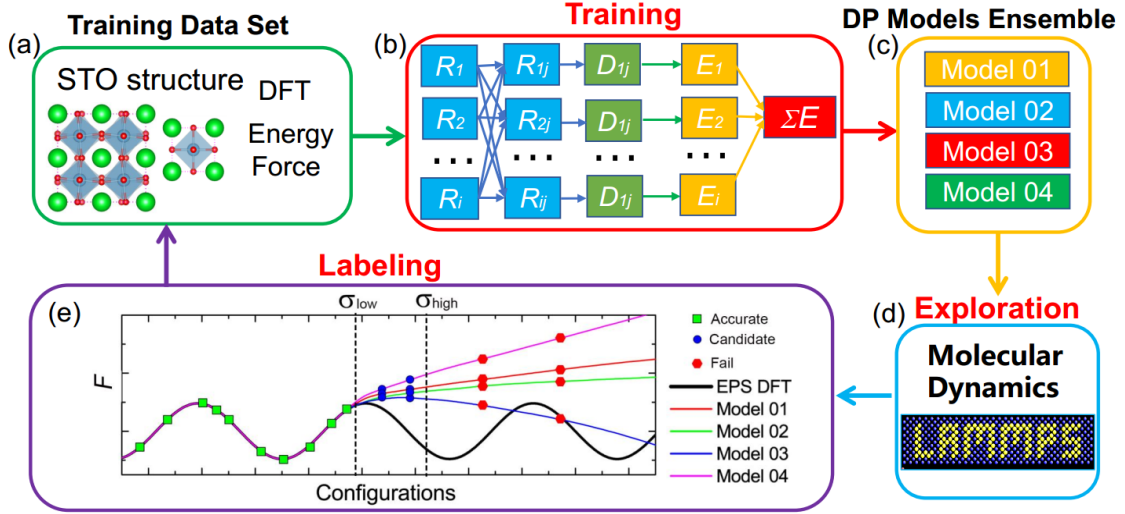

Fig.1 Schematic illustration of the DP-GEN process, including training, exploration, and labeling. (a) The training dataset contains hundreds of perturbed structures and corresponding DFT energies and forces. (b) Starting with these DFT training data set, four DP models (c) are trained starting with different values of deep neural network parameters. (d) Performing one of the DP models is used for MD simulations to explore the configuration space at different temperatures. (e) Labeling selected candidates by DFT calculation and then added to the training dataset of next iteration.

### B. Generation of the training data set

In this work, we use the deep potential generator (DP-GEN) to generate a set of training data that covers a sufficient wide range of relevant configurational space efficiently [39]. DP-GEN is a concurrent learning procedure and the workflow of each iteration includes three main steps: training, exploration, and labeling (see Fig. 1).

We start with DFT relaxed ground-state structures of *Pm-3m* cubic and *I4/mcm* tetragonal phase. The lattice constants calculated by DFT with Perdew-Burke-Ernzerhof solid (PEBsol) are shown in Table I. In the first iteration, the initial training dataset contains 200 randomly perturbed structures of $2 \times 2 \times 2$ *Pm-3m* cubic and $1 \times 1 \times 1$ *I4/mcm* tetragonal cell. Starting with these DFT training data set, four DP models are trained starting with different values of deep neural network parameters. Then, the exploration step was performed in which one of the DP models is used for MD simulations to explore the configuration space at different temperatures. For all sampled configurations in MD trajectories, the other three DP models will predict the corresponding energies and atomic forces. The maximum deviation of the four DP model prediction of forces ($\sigma_f^{max}$) can be used to formulate the criterion for labeling:

$$\sigma_f^{max} = \max_i \sqrt{\langle \|\boldsymbol{F}_i - \langle \boldsymbol{F}_i \rangle\|^2 \rangle},$$

where $\langle ... \rangle$ indicates the average of model predictions. The configurations with $\sigma_f^{max} < \sigma_{low}$ are already well represented by previous training dataset (green points in Fig. 1e), which are labeled as accurate configurations. Whereas a configuration with a large model deviation $\sigma_f^{max} > \sigma_{high}$ is highly distorted and unphysical due to poor model quality (red points in Fig. 1e), and is thus labeled as failed configuration. Therefore, only configuration satisfying $\sigma_{low} < \sigma_f^{max} < \sigma_{high}$ are labeled as candidates for further DFT calculations and are added to the training dataset for training in the next iteration (blue points in Fig. 1e). Here, $\sigma_{low}$ and $\sigma_{high}$ are set to 0.05 eV/Å and 0.20 eV/Å respectively as suggested in the original literature [39]. It is important to point out that $\sigma_f^{max}$ can be also used as the convergence criterion for DP-GEN iterations. Here, the exploration of each system is considered converged when the percentage of accurate configurations ($\sigma_f^{max} < \sigma_{low}$) is larger than 99 %. For more details of DP-GEN method, please refer to the original literature [22].

## C. DFT calculations setting

The initial training dataset is obtained by performing a 10-step *ab initio* MD simulation for randomly perturbed structures at 50 K. After labeling candidate configurations, self-consistent DFT should be performed. All DFT calculations were performed using a plane-wave basis set with a cutoff energy of 500 eV as implemented in the Vienna Ab initio Simulation Package (VASP) [43,44], the electron exchange-correlation potential was described using the generalized gradient approximation and PBEsol scheme [45]. The Brillouin zone was sampled with a 6 × 6 × 6 *k*-point grid for 2 × 2 × 2 *Pm-3m* supercell, 6 × 6 × 5 *k*-point grid for 1 × 1 × 1 *I4/mcm* cell.

## D. MD simulations setting

The MD simulations were carried out using LAMMPS code with periodic boundary conditions[46]. At the exploration step, MD simulations adopt the isobaric-isothermal (*NPT*) ensemble with temperature set from 50 to 400 K, pressure set from 0.001 to 50 kbar, and time with 2 to 20 ps (Table S1). The Nose–Hoover thermostat and Parrinello-

Rahman barostat are employed to control temperature and pressure, respectively [47,48]. After DP-GEN iteration is considered converged, the optimized DP models of STO can be used to study the structural transitions driven by temperature and strain via performing MD simulations. The *NPT* MD simulations starting with a ground-state tetragonal *I4/mcm* 10 × 10 × 10 pseudo-cubic supercell (5000 atoms) were performed with periodic boundary conditions and at the zero pressure to model the temperature and strain-driven structural transition. The time step is set to 0.001 ps. Simulations using 20 × 20 × 20 supercell give similar results. At a specified temperature, the equilibrium run is 10 ps, followed by a production run of 90 ps (see Fig. S2 [49]). The main results are not sensitive to the initial configuration (*Pm-3m* or *I4/mcm*) for MD simulation. To apply a strain of $\varepsilon_s$, the in-plane lattice constant of the supercell was fixed at $a_0+\varepsilon_s a_0$ Å, where $a_0$ is the optimized in-plane lattice constant of STO at the corresponding temperature, and the out-of-plane lattice constant was allowed to fully relax.

**E. DP-GEN iteration process**

The distributions of $\sigma_f^{max}$ at different temperatures in the first five iterations are shown in Fig. S1 [49], and the detailed percentage values of the accurate, candidate, and failed configurations in each iteration are shown in Table S1 [49]. In the first iteration, starting with two systems of 2 × 2 × 2 *Pm-3m* supercell (40 atoms) and 1 × 1 × 1 *I4mcm* cell (20 atoms), a lot of the configurations were labeled as candidate (39.5 %) and failed (16.8 %), because the preliminary DP model is very rough due to the limited information in the training dataset. After adding hundreds of candidate configurations into the training data set in the second iteration and a complete training, all sampled low-temperature configurations (50-200 K) were labeled as accurate and candidate configurations come from higher-temperature simulations. To explore the wider configuration space, the lager supercells are added at the later exploration steps. The whole DP-GEN iteration process is started with 400 initial configurations, then explored the three million configurations. Finally, the 2600 candidate configurations were selected for labeling and added into training dataset after a total number of 15 iterations. All settings of the exploration strategy are listed in Table S1 [49].

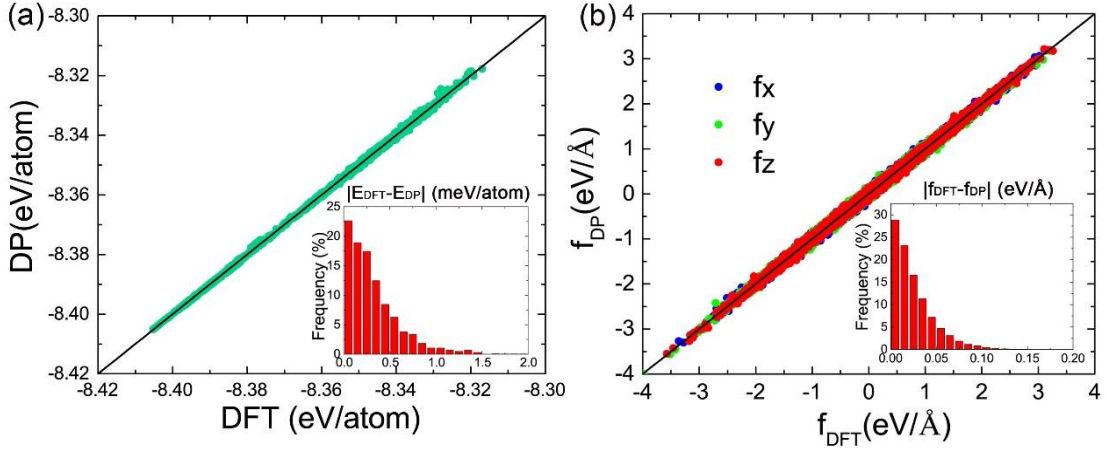

Fig.2 Comparison of (a) energies and (b) atomic forces calculated using the DP model and DFT calculation for the all configurations in the final training dataset.

## III. RESULTS AND DISCUSSIONS

### A. Accuracy of DP model

The whole deep learning process is parameter-free, which ensures that the generality and transferability of DP models. We compare the energies and atomic forces calculated using the DP model and DFT calculations for the configurations in the final training dataset. As shown in Fig. 2, the mean absolute error of energy ($\Delta E^{DP-DFT}$) and atomic force ($\Delta F^{DP-DFT}$) between DP and DFT is 0.322 meV/atoms and 0.025 eV/Å, respectively. The DFT and DP model energies of cubic *Pm-3m*, tetragonal *I4/mcm* phase are listed in Table I. For the structures of cubic and tetragonal phase relaxed by DP and DFT, the $\Delta E^{DP-DFT}$ is equal to 0.057 and 0.146 meV/atoms, respectively. The energy gain of a cubic phase relative to the ground-state tetragonal phase ($\Delta E^{C-T}$) is a key factor that determines the phase-transition temperature. The DFT calculated $\Delta E^{T-C}$ is 1.096 meV/atom, and DP calculated value is 1.008 meV/atom. It demonstrates that the DP model can capture a milli-electron-volts level energy different between ground-state tetragonal and cubic phase. It is also noted that the training dataset does not contain structural information of orthorhombic *Pnma* phase, but $\Delta E^{DP-DFT}$ for the *Pnma* STO is only 0.634 meV/atoms (Table I). It indicates that the DP model could well calculate the configurations which are not included in the training data set.

TABLE I. The DFT ($E^{DFT}$) and DP energies ($E^{DP}$) of *Pm-3m*, *I4/mcm*, and *Pnma* phase. $\Delta E^{T-C}$ is the energy of a *I4/mcm* phase relative to the ground-state *Pm-3m* phase.

| Structure | $E^{DFT}$ eV/atom | $E^{DP}$ eV/atom | Error meV/atom |
|---|---|---|---|
| *Pm3m* | -8.404051 | -8.403994 | 0.057 |
| *I4/mcm* | -8.405148 | -8.405002 | 0.146 |
| *Pnma* | -8.360826 | -8.361460 | 0.634 |
| $\Delta E^{T-C}$ | 0.001096 | 0.001008 | - |

TABLE II. Lattice constants of *Pm-3m* and *I4/mcm* phase at 0K predicted by DP model and DFT. $\varphi$ is the rotation angle of the TiO$_6$ octahedra around [001]

| Structure | Method | $a, b$ /Å | $c$ /Å | $\varphi$/° |
|---|---|---|---|---|
| *Pm-3m* | DFT | 3.895 | - | - |
| | DP | 3.898 | - | - |
| *I4/mcm* | DFT | 3.885 | 3.906 | 5.49 |
| | DP | 3.888 | 3.910 | 5.49 |

Table III. The elastic constants, bulk modulus ($Bv$), shear modulus ($Gv$), and Poisson's ratio ($v$) predicted by DP model and DFT. The unit of modulus is GPa.

|  | Cubic $Pm$-$3m$ | | $I4/mcm$ | |
|---|---|---|---|---|
|  | DFT | DP | DFT | DP |
| $Bv$ | 173 | 182.29 | 173 | 187.68 |
| $Gv$ | 110 | 100.89 | 109 | 102.91 |
| $v$ | 0.24 | 0.23 | 0.25 | 0.247 |
| $C_{11}$ | 351.93 | 363.18 | 352.23 | 337.752 |
| $C_{12}$ | 104.74 | 106.85 | 103.38 | 88.561 |
| $C_{44}$ | 116.28 | 100.89 | 115.99 | 99.766 |
| $C_{33}$ | - | - | 357.15 | 344 |
| $C_{13}$ | - | - | 107.87 | 123.01 |
| $C_{66}$ | - | - | 115.95 | 109.2 |

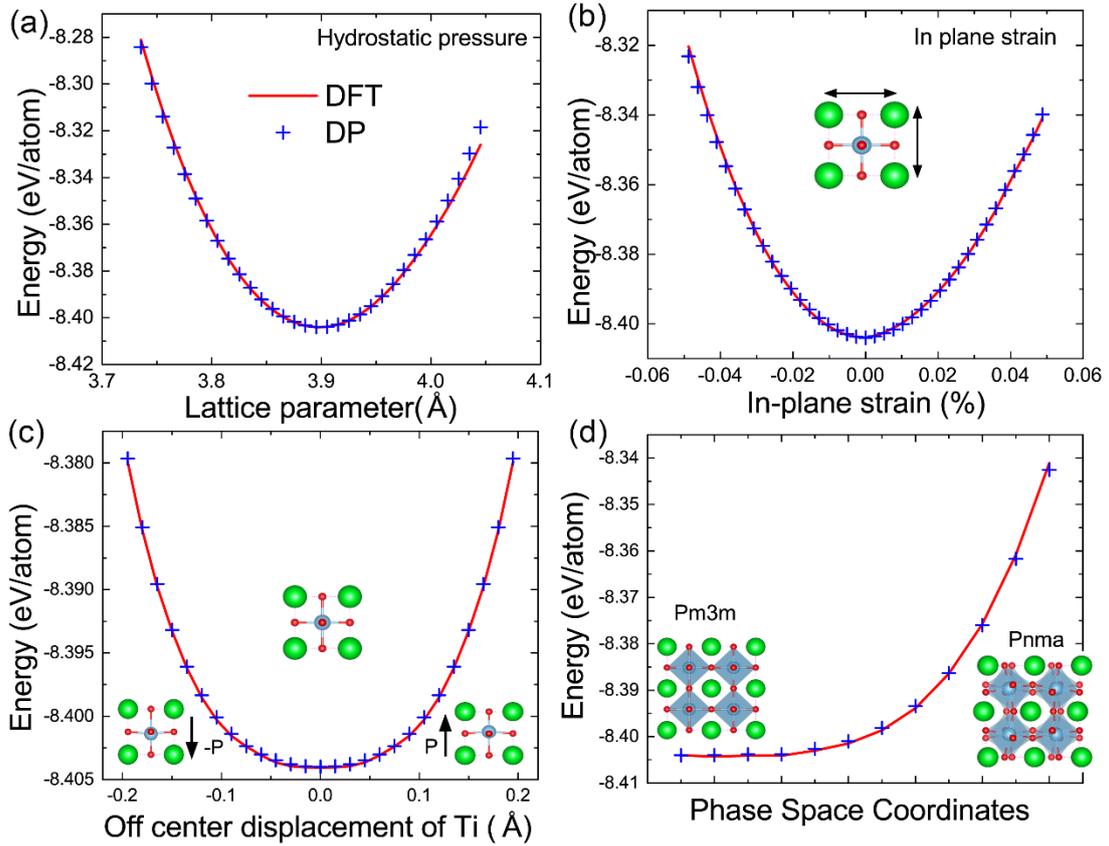

Fig.3 The DP predicted and DFT calculated energies for (a) hydrostatic pressure, (b) in plane strain, (c) Ti atom displacement of cubic phase, and (d) phase transition trajectory from cubic $Pm3m$ to orthorhombic $Pnma$ phase. It should be note that the $Pnma$ phase are not included in the training data set of the DP model.

Table II summarizes equilibrium lattice constants optimized by DP and DFT for two phases of STO at 0 K. It can be seen that the lattice constants for the DP model are within 0.004 Å of the DFT results. We also use the DP model to calculate the elastic constants for cubic and tetragonal phases, and compare the values with DFT results. The results are shown in Table III, demonstrating great agreement between DP and DFT. The equations of state of cubic phase calculated by DFT and DP model are presented in Fig. 3a. DP model reproduces well the DFT results over a wide range of lattice constants (3.7-4.0 Å). Adding more configurations of large lattice constants to the training dataset can further improve the DP model by fixing the slight deviation beyond 4.0 Å. Many experiments have demonstrated that the ferroelectric phase can be induced in STO via strain engineering [25,28,50]. We use DP model to predict the energy variation of cubic phases under in-plane biaxial strain and polarization (i.e. off-centering displacement of Ti atom). The results are shown in Fig. 3b and c, again demonstrating excellent agreement between DP and DFT results. We demonstrate that the DP model can accurately describe the orthorhombic *Pnma* phase STO the information of which was not included in the training dataset. We further explore the structure-energy relationship along the phase transition trajectory from *Pm3m* to *Pnma* phase by the DP model, and the DP results are in good agreement with DFT results as shown in Fig. 3d.

We also calculate the phonon dispersion relations and corresponding density of states of the cubic phase by the DP model and DFT. As shown in Fig. 4, DP model calculated phonon dispersion of the cubic STO agrees well with DFT spectrum, showing multiple lattice instabilities, i.e., an unstable antiferrodistortive (AFD) zone-boundary $R_5^-$ mode that corresponds to concerted rotations of the oxygen octahedra leading to the low-temperature tetragonal structure and an unstable zone-center polar $\Gamma_4^-$ mode associated with the incipient ferroelectricity. The slight difference of the polar $\Gamma_4^-$ mode at the Γ-point between DP and DFT could be eliminated if adding configurations with atoms perturbed according to the polar $\Gamma_4^-$ mode.

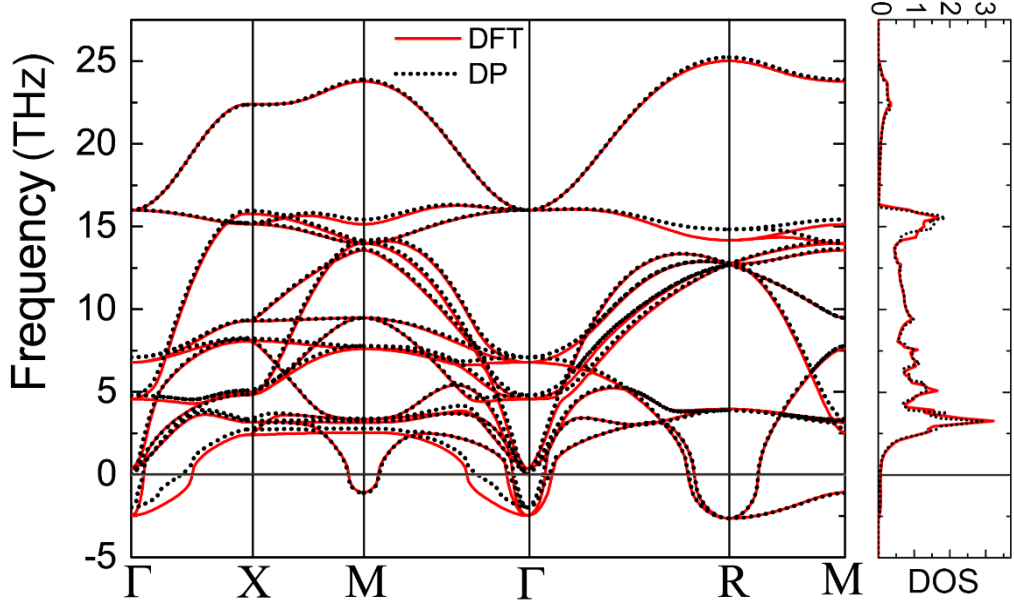

Fig. 4 The phonon dispersion relations and density of states of cubic phase of STO calculated by DP model and DFT.

## B. Temperature driven phase transition

Using the DP model, the atomic structure of STO at the finite temperature can be determined form DPMD simulations. Considering the AFD of staggered octahedra rotation around [001] in the tetragonal phase, we define $\varphi_n = (-1)^n \theta_n$ as the AFD order parameter, where the index $n$ is the sequence number of unit cell and $\theta$ is the rotation angle of TiO$_6$ octahedra in each unit cell. The temperature dependent lattice constants and $\varphi$ are shown in Fig.5. The simulated system has a tetragonal phase at the low temperature. As the temperature increases, both the tetragonal distortion ($c/a$) and $\varphi$ decrease. The $\varphi$ goes from a value of 4.61° at 50 K to 0.22° at 200 K. To get the atomistic insights into the phase transition, the distribution of $\varphi$ within each unit cell at different temperatures are shown in Fig. 6. It is indicated that the tetragonal phase has a long-range AFD ordering pattern at low temperature, and that the long-range feature vanishes at higher temperature where we observed short-range AFD ordering patterns appearing dynamically in space at 200 K (see Fig. 6c). According to the variation of lattice constant and $\varphi$, the transition from tetragonal to cubic phase takes place when the temperature is raised over 190 K. The DP model predicted behavior of temperature driven phase transition in STO is in qualitative agreement with experiment, but the transition temperature is slightly higher than the experimental value of 105 K [51]. In addition, DPMD predicts a linear behavior of *a* and *b*-axis lattice constants at low temperature, whereas the saturated temperature dependence of the lattice parameter below 50 K was observed in the experiment [52]. The overestimate of transition temperature and linear behavior of lattice constants at low temperature is likely a consequence of classical limitation of standard MD simulation. This is due to quantum effect plays an important role at low temperatures [14,15], but it is neglected in the standard MD simulations. Adding the statistical quantum features into MD could eliminate the disagreement between experiment and DPMD simulation [53-55].

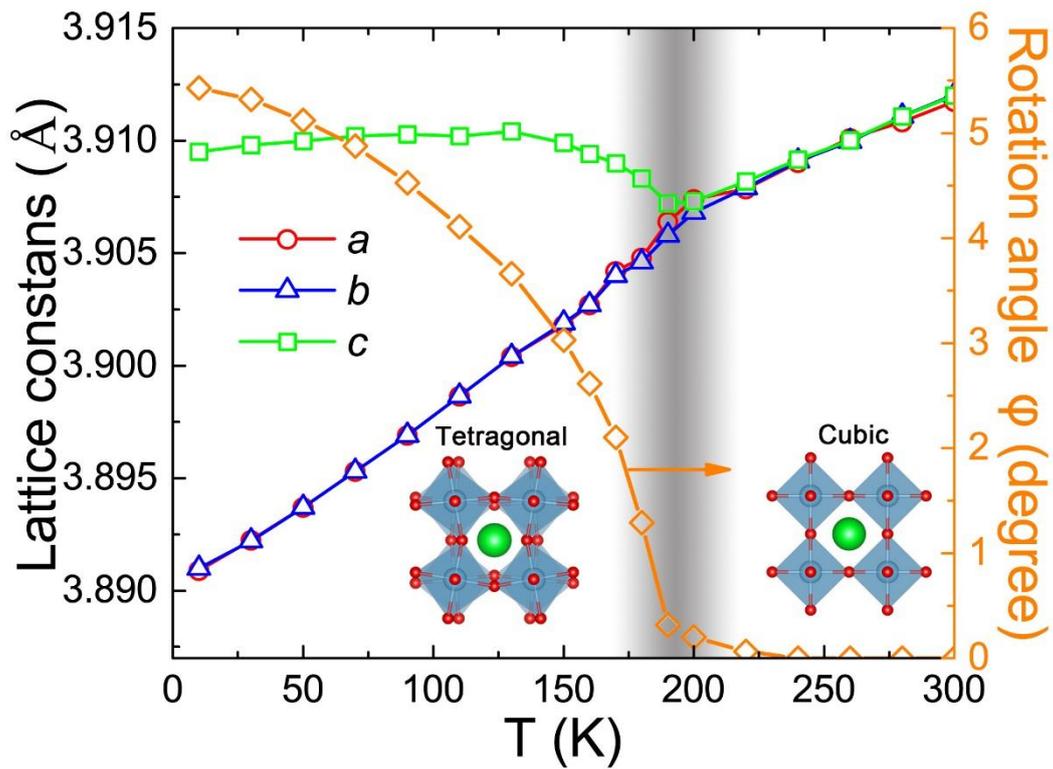

Fig. 5 Average lattice constants and rotation angle of the $TiO_6$ octahedra around [001] ($\varphi$) for the $SrTiO_3$ as a function of temperature determined from DPMD simulations with a 5000-atom supercell.

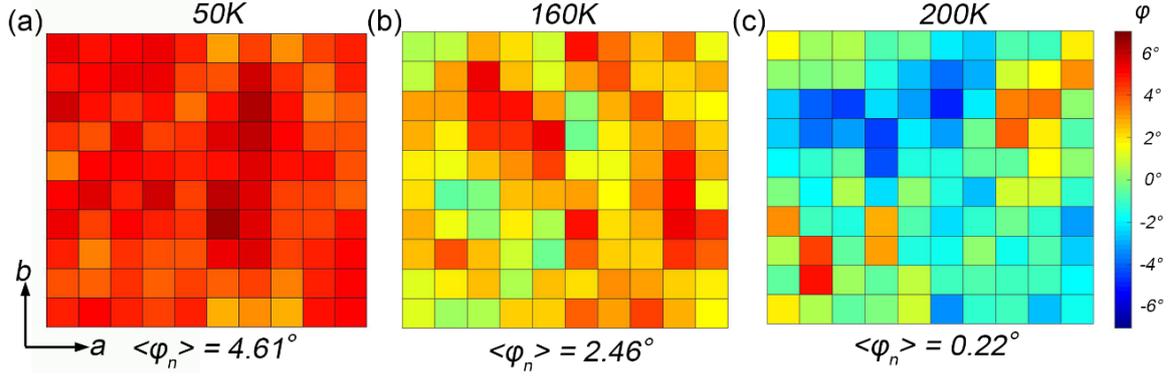

Fig. 6 The distribution of $\varphi$ within each unit cell in the *ab*-plane (in plane) at temperature of 30K (a), 160K (b), and 200K (c). The color illustrating the magnitude of $\varphi$. Considering the staggered octahedra rotation, the $\varphi$ is defined as $\varphi_n = (-1)^n \theta_n$, where the index $n$ is the sequence number of unit cell, $\theta$ is the rotation angle in each unit cell.

## C. Strain effect

It is well known that STO is an incipient ferroelectric so that the ferroelectric transition does not appear under strain-free state even when temperature decreases to 0 K. It is generally believed the ferroelectric order is suppressed by quantum fluctuation [14,15] and preceding AFD distortion [12,56]. No matter what type of the reasons it was, that the situation may change dramatically in the strained state. Many theoretical works and experiments reported epitaxial strain could induce a polar or ferroelectric phase in STO films[17,24-26,28,50]. Here, we predict the in-plane biaxial strain-temperature phase diagram of STO using DPMD simulations.

Fig. 7 shows the phase diagram of ferroelectric (FE) and AFD phase as a function of the temperature and in-plane biaxial strain for STO by DPMD. Here, we use two order parameters, the average polarization ($P$) and the AFD order parameter ($\varphi$), to distinguish the pure FE, FE, AFD, and paraelectric phase: the pure FE phase is defined as $P \neq 0$ and $\varphi = 0$, the FE phase is defined as $P \neq 0$ and $\varphi \neq 0$, the AFD phase is defined as $P = 0$ and $\varphi \neq 0$, whereas paraelectric phase is defined as $P = 0$ and $\varphi = 0$. It is seen that FE phase become most stable phase when the compressive (only when large than 0.2 %) and tensile strain was applied. At the low temperatures, the compressive strain can induce the out-of-plane $P$ and AFD structure with rotation axis along [001], whereas the tensile strain induces the in-plane $P$ and AFD structure with rotation axis along [100] and [010] (structural details see Fig. S3 and S4 [49]). Then the FE phase transfer to AFD structural phase when the temperature increases. It is clear to see that the ferroelectric transition temperature increases with the in-plane strain. When further increasing temperature, the AFD structure phase will transform to the paraelectric phase with $P$ and $\varphi$ both decreased to zero. The DP model shows that the ferroelectric transition temperature is ~110 K at 0.8 % compressive strain and ~210 K at 0.8 % tensile strain. DPMD simulations clarify that the wide region of possible ferroelectric transition range in the temperature-strain phase diagram predicted by LDG model is the AFD phase [11, 13]. In addition, it is interesting to note that the previously reported pure FE phase ($P \neq 0$ and $\varphi = 0$) was not observed in this our phase diagram [10, 11].

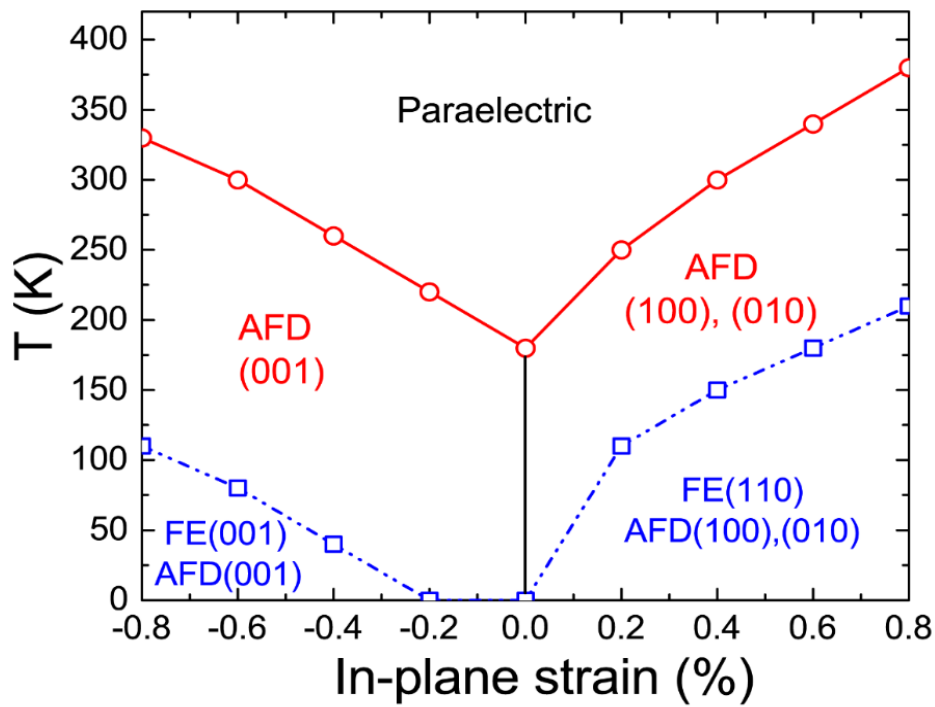

Fig. 7 Expected phase diagram of bulk SrTiO3 under biaxial in-plane strain by DPMD simulations. Ferroelectric and AFD transition temperature are plotted by blue square dot and red dot at the different strain.

We further investigate the nature of temperature-driven ferroelectric phase transition in compressively strained STO (strain of −0.8 %). We calculated the average $\varphi$ and $P$ as a function of temperature using DPMD simulations with results shown in

Fig.8a. It clearly reveals that the polarization and AFD structure coexist in the STO below the 110 K, and the polarization is induced by the off-centered displacement of Ti atoms and oxygen octahedron. The details of such atomic structure are shown in the Fig. S3[49]. Also, the DPMD simulations indicate that the polarization magnitude decreases with temperature and becomes zero above a ferroelectric phase transition temperature of 110 K. Similar to the polarization case, the AFD structure (i.e. $\varphi$) also disappears above 320 K. To get the atomistic detail characteristics of temperature-driven ferroelectric phase transition, we calculate the distribution of local polarization of *c*-axis component in STO supercell at different temperatures, and the results are shown in Fig. 8b. At 30 K, the distribution of local polarization in ferroelectric phase is exhibits a single peak is located near ~−0.14 C/m$^2$. The corresponding snapshot of dipole configurations in the *bc*-plane (out-of-plane) are shown in Fig. 8c, indicating long-range polarization order along *c*-axis in the ferroelectric phase. As the temperature increase, the peak shifts to the lower value, suggesting a displacive ferroelectric phase transition. When the temperature further rises to 90 and 110 K, another peak emerged at the opposite position, indicating the reversal of polarization vector (see dipole configurations of 90 and 110 K in Fig. 8c). It demonstrates that ferroelectric phase transition in STO under strain has both displacive and order-disorder characteristics. These results are in line with theoretical and latest experiment studies that ferroelectric phase transition in STO under strain has the mixture of displacive and order-disorder characteristics [17,57,58]. Similar characteristics of phase transition are also found in tensilely strained STO (see Fig. S5 [49]).

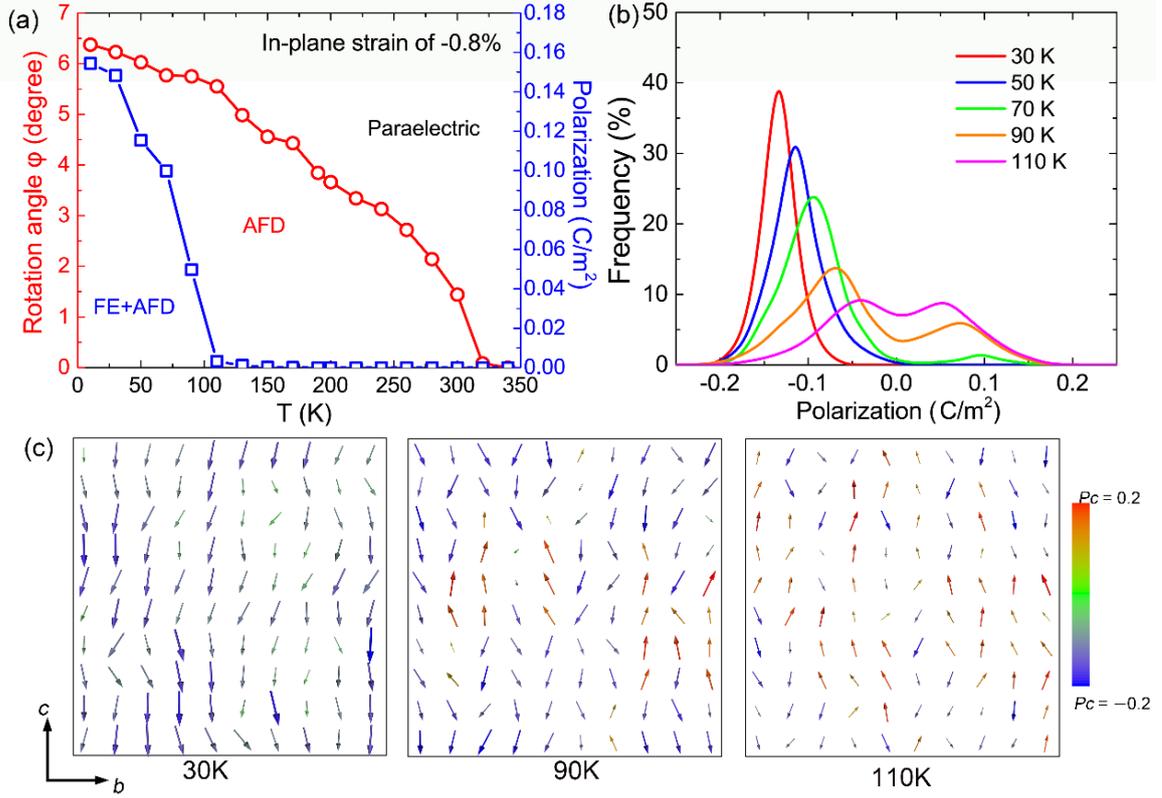

Fig. 8 Polarization (P) and rotation angle of the octahedra ($\varphi$) of 5000-atom supercell STO under -0.8% biaxial in-plane strain conditions. (a) P and $\varphi$ as a function of temperature. (b) The probability distribution of the local P in unit cell along the [001] direction at different temperatures. (c) Typical snapshots of dipole configurations in the *bc*-plane (out of plane) at different temperatures. Each arrow represents the local electric dipole vector within a pseudo-cubic unit cell.

## IV. CONCLUSIONS

In summary, we developed DP model to describe dynamic and energetic properties, and structural transition in STO with the accuracy of DFT at the PBEsol level. After confirming the accuracy of the DP models by comparison of lattice constant, free energy and other properties with DFT calculation results, we use the DP model to simulate temperature driven tetragonal to cubic phase transition at the finite temperature. Then, we predict the in-plane biaxial strain-temperature phase diagram of STO using DP model and get the atomistic insights into the ferroelectric and AFD structural transition. We clarify that ferroelectric (FE) phase at low temperature under strain is associated with AFD. With the temperature increasing, such FE phase transform into

AFD phase. Moreover, it is demonstrated that ferroelectric phase transition in STO under strain has characteristics of both displacive and order-disorder phase transition. These simulations provide new insights into the behavior of the ferroelectric phase transition of STO. Ultimately, the machine learning method provides significant promise for the successful development of interatomic potentials for more complex perovskite materials.

## ACKNOWLEDGMENTS

This work was supported by the National Key R&D Program of China (Grant No. 2017YFA0303602), the National Nature Science Foundation of China (Grants No. 11774360, No. 11974365, and No. 51931011), and Key Research Program of Frontier Sciences, CAS (Grant No. ZDBS-LY-SLH008).

**References**

[1] C. Zener, Phys. Rev. **82**, 403 (1951).
[2] G. Hodes, Science **342**, 317 (2013).
[3] D. D. Fong, G. B. Stephenson, S. K. Streiffer, J. A. Eastman, O. Auciello, P. H. Fuoss, and C. Thompson, Science **304**, 1650 (2004).
[4] A. Ohtomo and H. Hwang, Nature **427**, 423 (2004).
[5] Z. S. Popović, S. Satpathy, and R. M. Martin, Phys. Rev. Lett. **101**, 256801 (2008).
[6] A. F. Santander-Syro, O. Copie, T. Kondo, F. Fortuna, S. Pailhès, R. Weht, X. G. Qiu, F. Bertran, A. Nicolaou, A. Taleb-Ibrahimi *et al.*, Nature **469**, 189 (2011).
[7] A. D. Caviglia, M. Gabay, S. Gariglio, N. Reyren, C. Cancellieri, and J. M. Triscone, Phys. Rev. Lett. **104**, 126803 (2010).
[8] M. Ben Shalom, M. Sachs, D. Rakhmilevitch, A. Palevski, and Y. Dagan, Phys. Rev. Lett. **104**, 126802 (2010).
[9] Z. Zhong, A. Tóth, and K. Held, Phys. Rev. B **87**, 161102 (2013).
[10] Q.-Y. Wang, Z. Li, W.-H. Zhang, Z.-C. Zhang, J.-S. Zhang, W. Li, H. Ding, Y.-B. Ou, P. Deng, K. Chang *et al.*, Chin. Phys. Lett. **29**, 037402 (2012).
[11] J.-F. Ge, Z.-L. Liu, C. Liu, C.-L. Gao, D. Qian, Q.-K. Xue, Y. Liu, and J.-F. Jia, Nat. Mater. **14**, 285 (2015).
[12] W. Zhong and D. Vanderbilt, Phys. Rev. Lett. **74**, 2587 (1995).
[13] P. Fleury, J. Scott, and J. Worlock, Phys. Rev. Lett. **21**, 16 (1968).
[14] K. A. Müller and H. Burkard, Phys. Rev. B **19**, 3593 (1979).
[15] W. Zhong and D. Vanderbilt, Phys. Rev. B **53**, 5047 (1996).
[16] J. H. Haeni, P. Irvin, W. Chang, R. Uecker, P. Reiche, Y. L. Li, S. Choudhury, W. Tian, M. E. Hawley, B. Craigo *et al.*, Nature **430**, 758 (2004).

[17] R. Xu, J. Huang, E. S. Barnard, S. S. Hong, P. Singh, E. K. Wong, T. Jansen, V. Harbola, J. Xiao, and B. Y. Wang, Nat. Commun. **11**, 1 (2020).

[18] R. He, J. L. Lin, Q. Liu, Z. Liao, L. Shui, Z. J. Wang, Z. Zhong, and R.-W. Li, ACS Appl. Mater. Interfaces **12**, 45602 (2020).

[19] D. Lee, H. Lu, Y. Gu, S.-Y. Choi, S.-D. Li, S. Ryu, T. R. Paudel, K. Song, E. Mikheev, S. Lee *et al.*, Science **349**, 1314 (2015).

[20] T. Mitsui and W. B. Westphal, Phys. Rev. **124**, 1354 (1961).

[21] T. F. Nova, A. S. Disa, M. Fechner, and A. Cavalleri, Science **364**, 1075 (2019).

[22] X. Li, T. Qiu, J. Zhang, E. Baldini, J. Lu, A. M. Rappe, and K. A. Nelson, Science **364**, 1079 (2019).

[23] S. Hayward and E. Salje, Phase Transit. **68**, 501 (1999).

[24] N. Pertsev, A. Tagantsev, and N. Setter, Phys. Rev. B **61**, R825 (2000).

[25] Y. Li, S. Choudhury, J. Haeni, M. Biegalski, A. Vasudevarao, A. Sharan, H. Ma, J. Levy, V. Gopalan, and S. Trolier-McKinstry, Phys. Rev. B **73**, 184112 (2006).

[26] H. Uwe and T. Sakudo, Phys. Rev. B **13**, 271 (1976).

[27] R. Wahl, D. Vogtenhuber, and G. Kresse, Phys. Rev. B **78**, 104116 (2008).

[28] S. E. Reyes-Lillo, K. M. Rabe, and J. B. Neaton, Phys. Rev. Mater. **3**, 030601 (2019).

[29] R. B. Wexler, Y. Qi, and A. M. Rappe, Phys. Rev. B **100**, 174109 (2019).

[30] S. Tinte, M. Stachiotti, S. Phillpot, M. Sepliarsky, D. Wolf, and R. Migoni, J. Phys.-Condes. Matter **16**, 3495 (2004).

[31] T. Katsumata, Y. Inaguma, M. Itoh, and K. Kawamura, Solid State Ionics **108**, 175 (1998).

[32] J. Behler and M. Parrinello, Phys. Rev. Lett. **98**, 146401 (2007).

[33] A. P. Bartók, M. C. Payne, R. Kondor, and G. Csányi, Phys. Rev. Lett. **104**, 136403 (2010).

[34] Y. Zhao, J. Fan, L. Su, T. Song, S. Wang, and C. Qiao, in *2020 IEEE 40th International Conference on Distributed Computing Systems (ICDCS)*2020), pp. 584.

[35] K. T. Schütt, H. E. Sauceda, P.-J. Kindermans, A. Tkatchenko, and K.-R. Müller, J. Chem. Phys. **148**, 241722 (2018).

[36] L. Zhang, J. Han, H. Wang, R. Car, and E. Weinan, Phys. Rev. Lett. **120**, 143001 (2018).

[37] L. Zhang, J. Han, H. Wang, W. A. Saidi, R. Car, and E. Weinan, in Proceedings of the 32nd International Conference on Neural Information Processing Systems, NIPS'18 (Curran Associates, Inc., Red Hook, NY, 2018), pp. 4441–4451.

[38] J. Wu, Y. Zhang, L. Zhang, and S. Liu, Phys. Rev. B **103**, 024108 (2021).

[39] Y. Zhang, H. Wang, W. Chen, J. Zeng, L. Zhang, H. Wang, and W. E, Comput. Phys. Commun. **253**, 107206 (2020).

[40] J. Wu, L. Bai, J. Huang, L. Ma, J. Liu, and S. Liu, Phys. Rev. B **104**, 174107 (2021).

[41] J. Huang, L. Zhang, H. Wang, J. Zhao, J. Cheng, and W. E, J. Chem. Phys. **154**, 094703 (2021).

[42] L. Zhang, D.-Y. Lin, H. Wang, R. Car, and W. E, Phys. Rev. Mater. **3**, 023804 (2019).

[43] G. Kresse and J. Furthmüller, Comput. Mater. Sci. **6**, 15 (1996).

[44] G. Kresse and J. Furthmüller, Phys. Rev. B **54**, 11169 (1996).

[45] J. P. Perdew, K. Burke, and M. Ernzerhof, Phys. Rev. Lett. **78**, 1396 (1997).

[46] S. Plimpton, J. Comput. Phys. **117**, 1 (1995).

[47] M. Parrinello and A. Rahman, J. Appl. Phys. **52**, 7182 (1981).

[48] S. Nosé, J. Chem. Phys. **81**, 511 (1984).

[49] See Supplemental Material at http://link.aps.org/supplemental/10.1103/PhysRevB.xx.xxxxxx

for Tables S1, and Figs. S1-S5.

[50] J. Haeni, P. Irvin, W. Chang, R. Uecker, P. Reiche, Y. Li, S. Choudhury, W. Tian, M. Hawley, and B. Craigo, Nature **430**, 758 (2004).

[51] T. Riste, E. J. Samuelsen, K. Otnes, and J. Feder, Solid State Commun. **9**, 1455 (1971).

[52] A. Okazaki and M. Kawaminami, Mater. Res. Bull. **8**, 545 (1973).

[53] H. Dammak, Y. Chalopin, M. Laroche, M. Hayoun, and J.-J. Greffet, Phys. Rev. Lett. **103**, 190601 (2009).

[54] J.-L. Barrat and D. Rodney, J. Stat. Phys. **144**, 679 (2011).

[55] H. Dammak, M. Hayoun, F. Brieuc, and G. Geneste, Journal of Physics: Conference Series **1136**, 012014 (2018).

[56] H. Vogt, Phys. Rev. B **51**, 8046 (1995).

[57] A. Bussmann-Holder, H. Beige, and G. Völkel, Phys. Rev. B **79**, 184111 (2009).

[58] S. Salmani-Rezaie, K. Ahadi, W. M. Strickland, and S. Stemmer, Phys. Rev. Lett. **125**, 087601 (2020).

# Supplemental Material

# Structural Phase Transitions in SrTiO$_3$ from Deep Potential Molecular Dynamics

Ri He[1#], Hongyu Wu[1#], Linfeng Zhang[2,3], Xiaoxu Wang[2,3], Fangjia Fu[3,4], Shi Liu[5,6,7*] and Zhicheng Zhong[1,8*]

[1]Key Laboratory of Magnetic Materials Devices & Zhejiang Province Key Laboratory of Magnetic Materials and Application Technology, Ningbo Institute of Materials Technology and Engineering, Chinese Academy of Sciences, Ningbo 315201, China

[2]DP Technology, Beijing 100080, China

[3]AI for Science Institute, Beijing 100080, China

[4]School of mathematical sciences, Peking University, Beijing 100871, China

[5]School of Science, Westlake University, Hangzhou, Zhejiang 310024, China

[6]Institute of Natural Sciences, Westlake Institute for Advanced Study, Hangzhou, Zhejiang 310024, China

[7]Key Laboratory for Quantum Materials of Zhejiang Province, Hangzhou Zhejiang 310024, China

[8]China Center of Materials Science and Optoelectronics Engineering, University of Chinese Academy of Sciences, Beijing 100049, China

---

[#] These authors contribute equally to this work
[*] liushi@westlake.edu.cn
[*] zhong@nimte.ac.cn

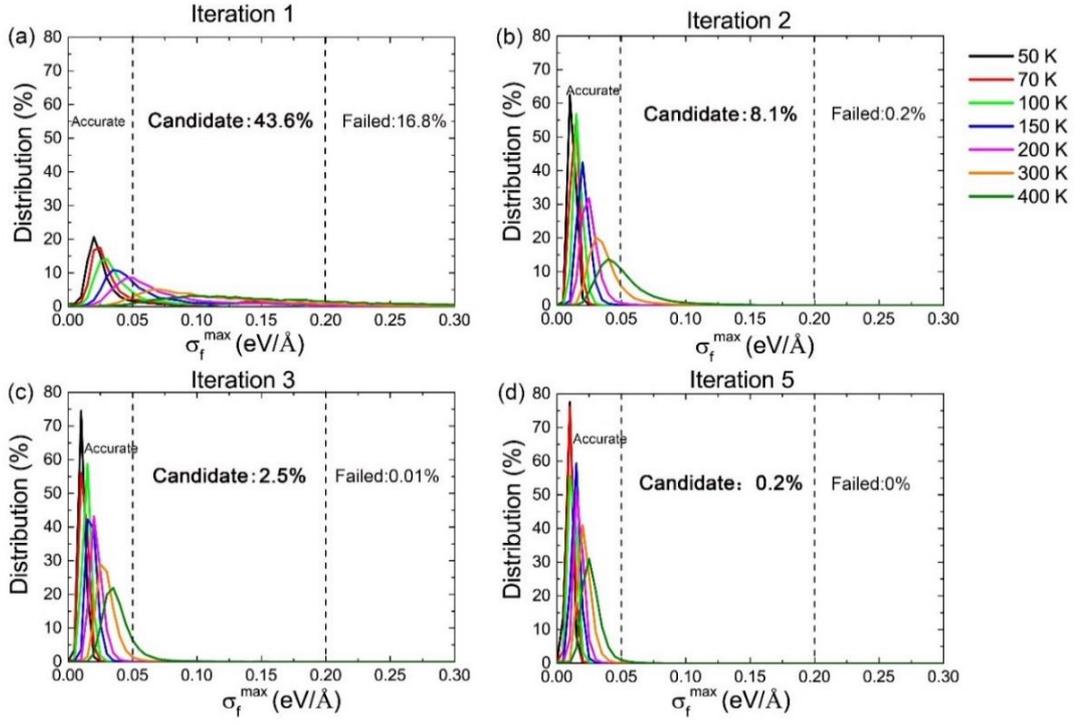

Fig. S1 The distribution of $\sigma_f^{max}$ at different temperatures in the iteration 1-5.

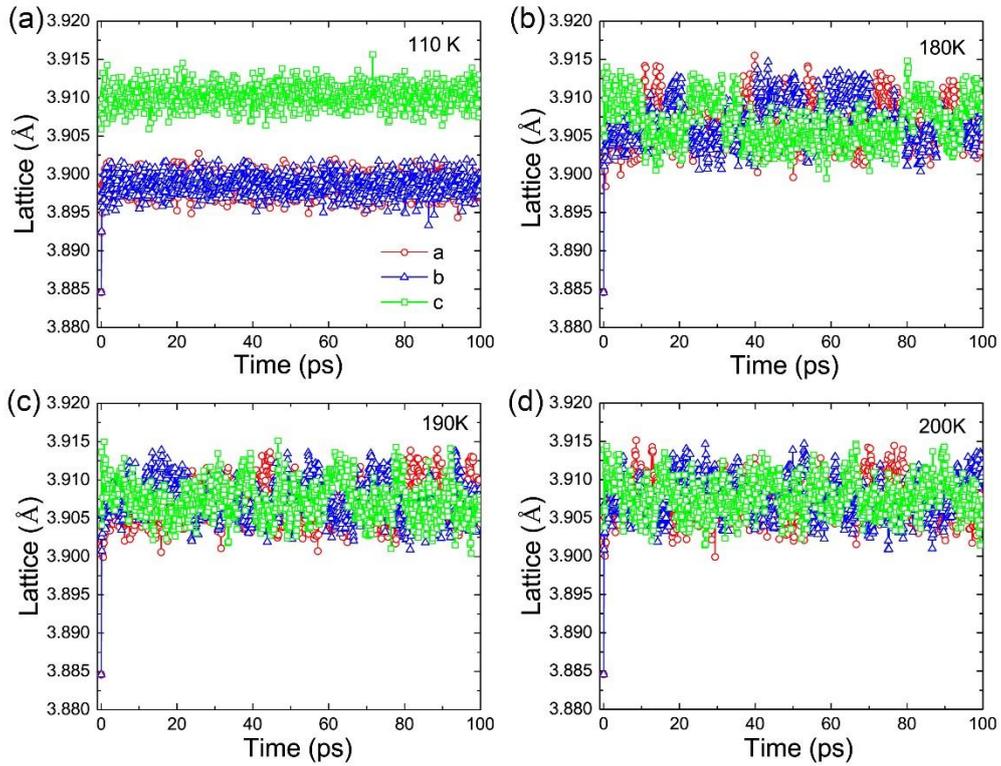

Fig. S2 Evolution of lattice constants of SrTiO$_3$ as a function of simulation time with a 5000-atom supercell.

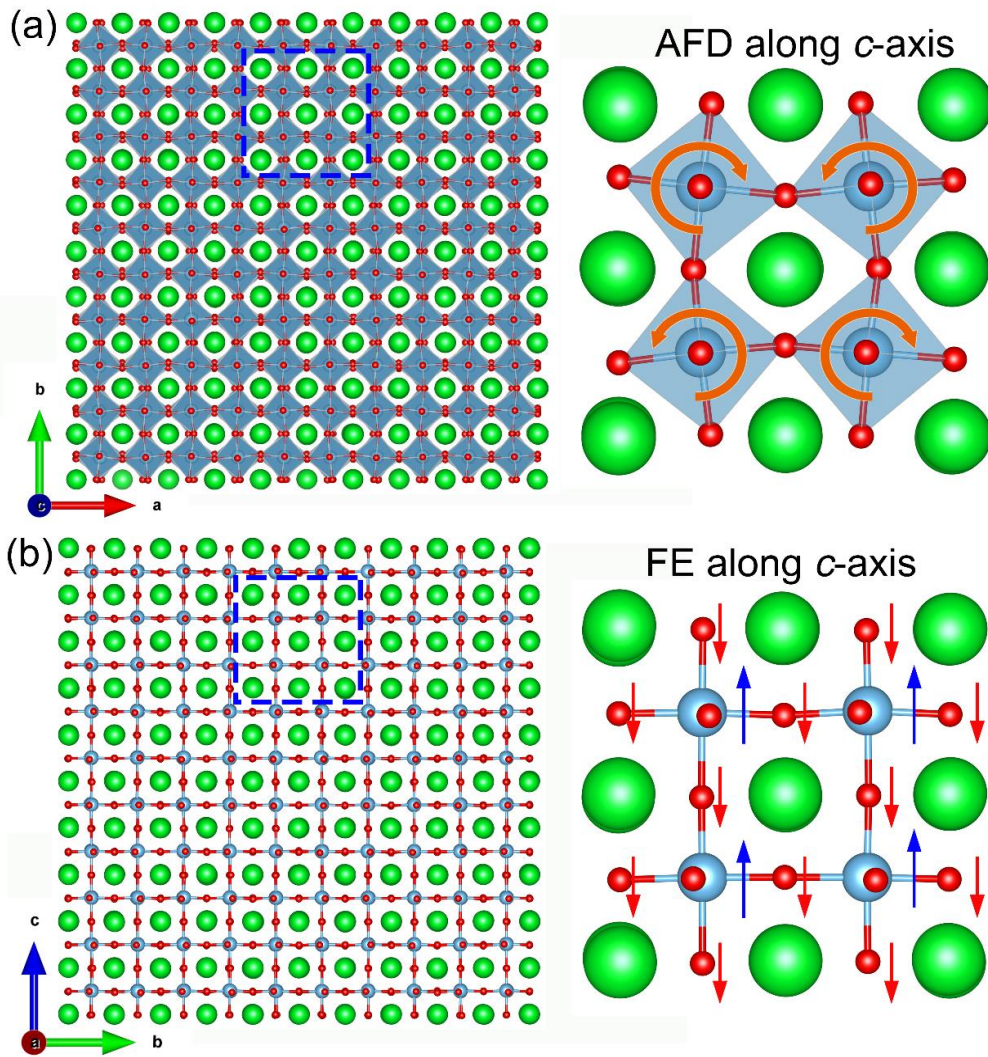

Fig. S3 Compressive in-plane strain induces the out-of-plane polarization and AFD mixed structure of STO.

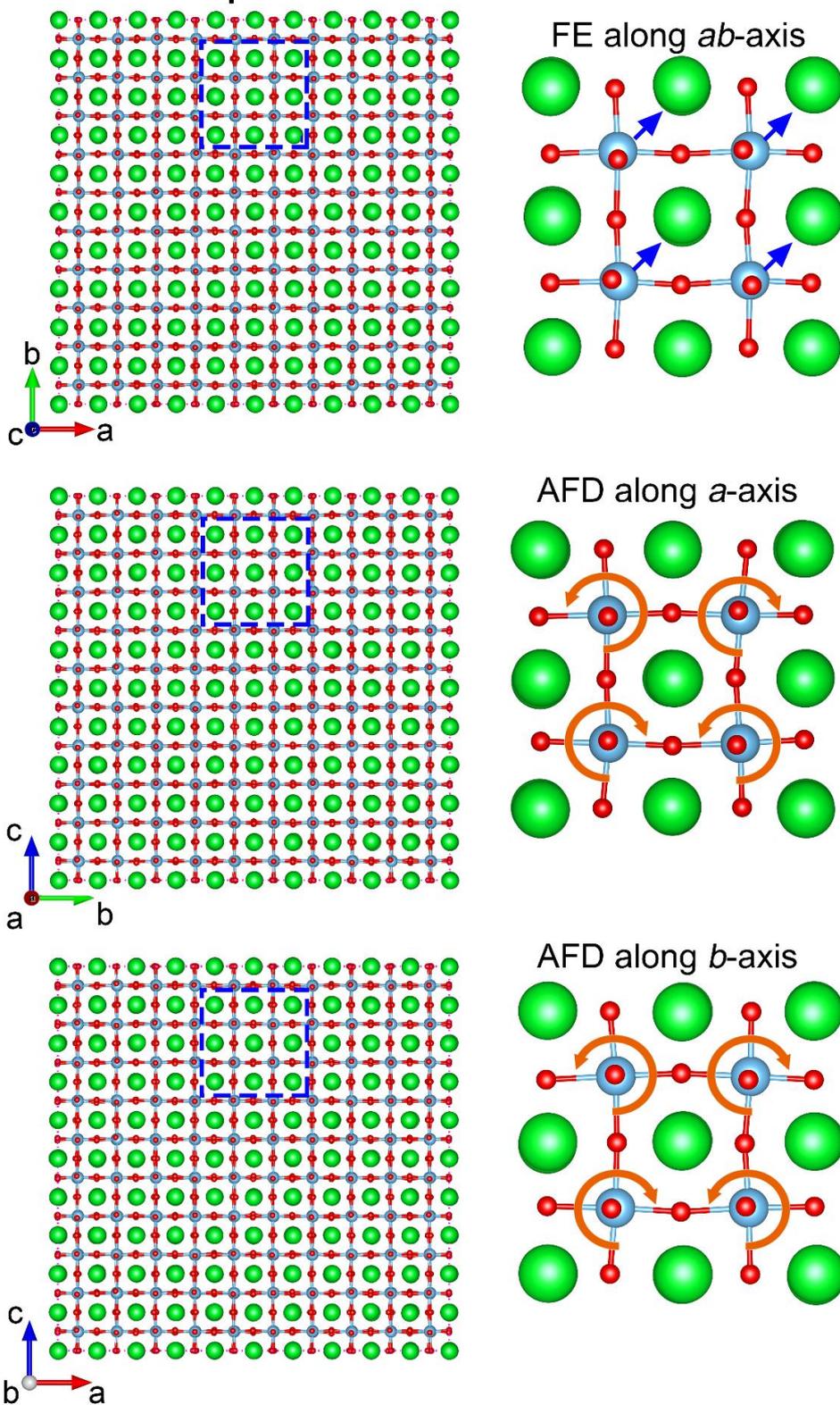

Fig. S4 Tensile in-plane strain induces the in-plane polarization and AFD mixed structure.

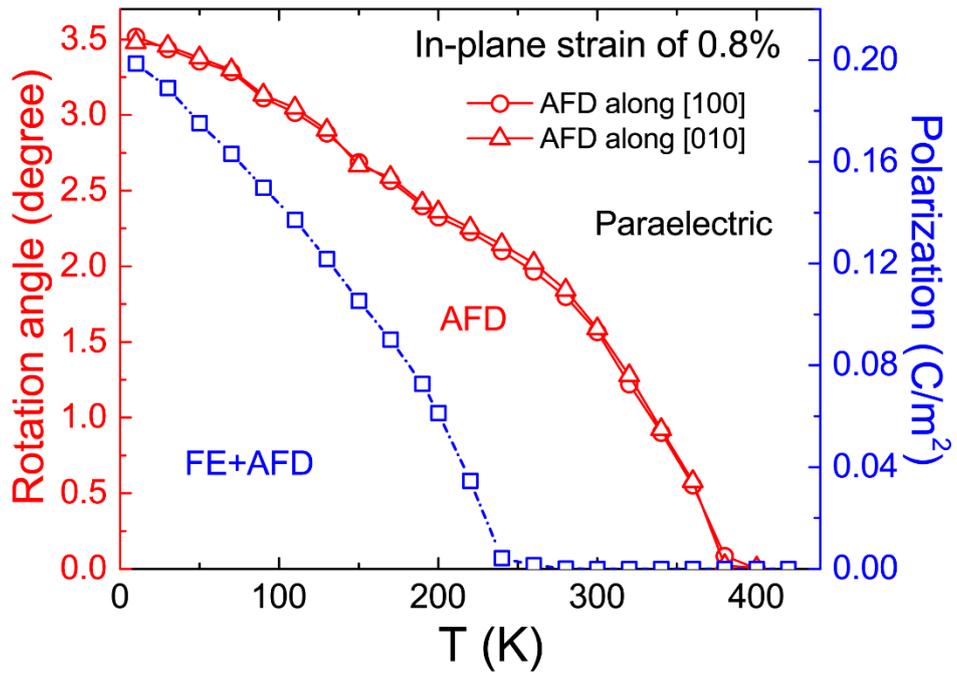

Fig. S5 Polarization and AFD order parameter $\varphi$ as a function of temperature under 0.8% biaxial in-plane strain conditions.

Table S1. Exploration settings of DP-GEN iterations.

| Iter. | Structures Supercell | Length (ps) | P (bar) | T (K) | Candidate % | Accurate % | Failed % |
|---|---|---|---|---|---|---|---|
| 1 | Pm-3m 2×2×2 (40 atom) I4/mcm 1×1×1 (20 atom) | 2 | 1, 10, 100, 1000, 10000, 20000, 50000 | 50, 70, 100, 150, 200, 300, 400 | 43.63 | 39.52 | 16.86 |
| 2 | | 6 | | | 8.07 | 91.74 | 0.2 |
| 3 | | 10 | | | 2.51 | 97.48 | 0.01 |
| 4 | | 16 | | | 0.86 | 97.15 | 1.99 |
| 5 | | 20 | | | 0.23 | 99.76 | 0 |
| 6 | Pm-3m 2×2×2 (40 atom) I4/mcm 1×1×1 (20 atom) I4/mcm 2×1×1 (40 atom) | 2 | 1, 10, 100, 1000, 10000, 20000, 50000 | 50, 70, 100, 150, 200, 300, 400 | 22.55 | 77.45 | 0 |
| 7 | | 6 | | | 0.25 | 99.75 | 0 |
| 8 | | 10 | | | 0.85 | 99.14 | 0 |
| 9 | | 16 | | | 0.20 | 99.80 | 0 |
| 10 | | 20 | | | 0.12 | 99.88 | 0 |
| 11 | Pm-3m 2×2×2 (40 atom) I4/mcm 1×1×1 (20 atom) I4/mcm 2×1×1 (40 atom) I4/mcm 2×2×1 (80 atom) | 2 | 1, 10, 100, 1000, 10000, 20000, 50000 | 50, 70, 100, 150, 200, 300, 400 | 0.95 | 99.05 | 0 |
| 12 | | 6 | | | 1.4 | 98.6 | 0 |
| 13 | | 10 | | | 1.09 | 98.91 | 0 |
| 14 | | 16 | | | 4.44 | 95.56 | 0 |
| 15 | | 20 | | | 1.24 | 98.76 | 0 |